\documentclass[twocolumn,showpacs,preprintnumbers,amsmath,amssymb]{revtex4}

\usepackage{graphicx}
\usepackage{dcolumn}
\usepackage{bm}

\begin{document}

\preprint{APS/123-QED}

\title{Abrupt Grain Boundary Melting in Ice}

\author{L. Benatov$^{1}$ and J.S. Wettlaufer$^{1,2}$}
\affiliation{Departments of Geology and Geophysics$^{1}$ and Physics$^{2}$\\Yale University, New Haven, Connecticut 06520}

\date{\today}

\begin{abstract}
The effect of impurities on the grain boundary melting of ice is
investigated through an extension of Derjaguin-Landau-Verwey-Overbeek
theory, in which we include retarded potential effects in a
calculation of the full frequency dependent van der Waals and
Coulombic interactions within a grain boundary. At high dopant
concentrations the classical solutal effect dominates the melting
behavior.  However, depending
on the amount of impurity and the surface charge density,  as
temperature decreases, the attractive tail of the dispersion force
interaction begins to compete effectively with the repulsive screened
Coulomb interaction.  This leads to a film-thickness/temperature curve that
changes depending on the relative strengths of these interactions and exhibits a {\em decrease} in the film thickness with {\em increasing} impurity level. More striking is
the fact that at very large film thicknesses, the repulsive Coulomb
interaction can be effectively screened leading to an abrupt
reduction to zero film thickness.

\end{abstract}


\maketitle

\section{\label{sec:level1}Introduction}

The original microscopic theories of melting focused on the criterion for the homogeneous breakdown of a harmonic solid lattice \cite{Dash02}.  Elaborations of this perspective involve the thermal activation of defects or the instability of the lattice \cite{Polturak} but a theory of melting must treat bulk solid-liquid coexistence and recognize that solid matter in laboratory and natural conditions is finite and hence its bulk is disjoined from the surroundings by surfaces \cite{Dash02}, \cite{Polturak}.  
Hence, the melting of a {\it finite} crystal begins at one of its free surfaces at temperatures
below the bulk transition \cite{Dash02}.
This {\it interfacial premelting} occurs at the surfaces of solid
rare gases, quantum solids, metals, semiconductors and molecular
solids and is characterized by the appearance of an interfacial thin
film of liquid that grows in thickness as the bulk melting
temperature, $T_m$, is approached from below.  The relationship
between the film thickness and the temperature depends on the nature
of the interactions in the system.  When interfacial premelting
occurs at vapor surfaces it is referred to as {\it surface melting}
and when it occurs at the interface between a solid and a chemically
inert substrate it is called {\it interfacial melting}.  When films
at such solid surfaces diverge at the bulk transition the melting is
{\it complete} but where retarded potential effects intervene and
attentuate the intermolecular wetting forces the film growth may be
blocked and thereby be finite at the bulk transition.  This latter
circumstance, in which the behavior is discontinuous, is referred to
as {\it incomplete} melting. The importance of grain boundary melting
is great because of its potentially central influence on the
sintering, coarsening, transport behavior and many other bulk
properties in ostensibly all materials.  This fact has certainly not
gone unrecognized, but nonetheless, the great difficulty of directly
accessing a grain boundary in thermodynamic equilibrium has resulted
in a dearth of experimental tests.

It is observed that, at
temperatures below $T_m$, polycrystalline matter is threaded by the
liquid phase driven solely by the impurity and curvature depressions
of the freezing point (e.g., \cite{Veins} for ice and
\cite{Smith} for metals).   For example, the unfrozen water is found in microscopic
channels, with a scale of 10-100 $\mu$m, called veins where three
grains abut and at nodes separating four grains rather like the
plateau borders in foams \cite{Veins}. Hence, the equilibrium
structure of polycrystalline ice is characterized by a connected
network of water.  Nye and Frank \cite{NyeFrank} predicted the
geometry of the network under the assumption that the interfacial
energies are independent of crystallographic orientation. In exact
analogy with the classical force balance used to determine the
contact angle of a partially wetting fluid on a substrate or the meniscus in capillary rise (e.g.,
\cite{DeGennes}), their result considers the force balance at the
trijunction where three grains come together. This provides an
expression for the dihedral angle,  2$\theta_0$, into
which water intrudes, in terms of the simple ratio of the grain
boundary, $\gamma_{ss}$ , and solid-liquid,  $\gamma_{s\ell}$,
interfacial energies:
\begin{equation}
\label{dihedral}
2\cos{\theta_0} = \frac{\gamma_{ss}}{\gamma_{s\ell}},
\end{equation}
thereby determining the shape and cross section of a vein.
Therefore,  the concept of interfacial premelting and the vein-node
network are related through grain boundary melting.  Thus,  during
{\it complete} grain boundary melting the dihedral angle vanishes and
the single grain boundary is replaced by two interfaces; $\gamma_{ss}
\rightarrow 2\gamma_{s\ell}$ .  This process is equivalent to the
growth of a liquid film at the grain boundary and hence connects the
dihedral angle, the vein/node/trijunction network and grain boundary
melting .

Because most materials exist in a polycrystalline state,
their continuum properties, ranging from the mechanical and thermal
properties of glaciers \cite{Veins},\cite{NyeFrank} to the reduction in the critical current density 
in high temperature superconductors \cite{thompson},
are to a large extent controlled by the
character of grain boundaries.   However, although direct
experimental probes of surface and interfacially melted films in
thermodynamic equilibrium are common, experimental studies of
grain boundary melting are centered on dihedral angle measurements.
The disparity in the relevant length scales of grain boundary films
and the dihedral angle is such that while optical microscopy may be
sufficient to determine the latter, the former requires a more highly
resolved probe.  Grain boundary melting poses serious experimental
challenges because of the difficulty of direct access to the
equilibrium interface of a bicrystal in a manner free of compromises
associated with the proximity of surfaces associated with apparatus.
Although computer simulations and theory support the notion of
disorder at a grain boundary \cite{Kikuchi}-\cite{Lobkovsky},
experimentally, it is often stimulated using impurities and then
quenching a system before probing the boundary (e.g., \cite{Yet},
\cite{Roger}).  Experiments in aluminum using electron microscopy have
shown that as the temperature rises, the boundary structure remains
epitaxial until $T = T_m$ when the grain boundary has the signature
of bulk liquid \cite{Balluffi}.   Dihedral angle experiments on
bismuth bicrystals placed in contact with the melt were performed for
various crystallographic mismatches by Glicksman and colleagues (See
\cite{Marty} and refs therein).  They interpreted their data to indicate a discontinuous grain boundary
melting transition as a function of grain mismatch. Relative to the abundant
work on surface and interfacial melting, the present experimental edifice for grain
boundary melting in all materials is lacking.  From the standpoint of
materials science, the advantage  of studying ice is that the melting
temperature is easily accessible in the laboratory without
substantial cryogenics or heaters.  Moreover, ice is transparent and
birefringent and hence amenable to optical probing.  Therefore, the
dual importance of grain boundary melting in materials in general and
in ice in particular make a compelling case for systematic
measurements in ice.   Ice provides the ideal test bed for such
measurements and its propitious geophysical importance is a prime
motivator \cite{RPP}.  The theory described here is intended to form
a rostrum for experimental studies.

\section{\label{sec:level1}Theory}

\subsection{\label{Background}Background}

Complete interfacial melting is determined by the competition between bulk and surface free energies and requires that the total excess surface free energy per unit area, $F_{\rm total}(d)$, be a positive monotonically decreasing function of the film thickness with a global minimum at infinite film thickness.  In grain boundary melting $F_{\rm total}(d)$=$\gamma_{ss}(d)$.  It is conceptually useful to consider the thickness dependent contributions to  $\gamma_{ss}(d)$ to arise from ostensibly short-- and long--ranged intermolecular interactions; wetting forces.  Hence, we write the excess surface free energy as $\gamma_{ss}(d)= 2\gamma_{s{\ell}} + F_{\rm short}(d) +  F_{\rm long}(d)$, where as above $\gamma_{s{\ell}}$ is the interfacial free energy per unit area of the ice--water interfaces, with implicit reference to the crystallographic orientation present at an interface.   Hence, by definition, at large enough distances, the long range interactions  always dominate over  shorter range interactions.  Therefore, if interactions underlying $F_{\rm long}(d)$ are attractive, that is they are represented by a negative monotonically increasing function of $d$, then $\gamma_{ss}(d)$ can never have a global minimum at $d= \infty$.  However,  if a short range interaction $F_{\rm short}(d)$ favored grain boundary melting,  it may prevail over $F_{\rm long}(d)$ out to large film thicknesses;
$|F_{\rm short}(d^{\prime})|~\geq~| F_{\rm long}(d^{\prime})|$ for $0~\leq~d~\leq~{d^{\prime}}$.  The {\it magnitude} of the global
minimum in $\gamma_{ss}(d)$ may then be quite small but will nonetheless occur at
a large (but finite) value of $d\equiv d_{\rm min}$, that is,  $\gamma_{ss}(d_{\rm min}) \equiv {\rm min}\left[\gamma_{ss}(d)\right] \equiv {\Gamma}$.  
If this were the case, most physical
observations of the system would be virtually indistinguishable from those
expected for complete melting.  A complete calculation of
$\gamma_{ss}(d)$ allows one to estimate both the dihedral angle and the
temperature at which the film thickness would saturate; each as a function of the thickness $d_{\rm min}$,
The dihedral angle $\theta_0$ is
given by $\cos{\theta_0}=1+ \frac{{\Gamma}}{2 \gamma_{s\ell}}$, and hence, in the pure case, we estimate the temperature at which the film thickness saturates as
given by $T-T_m \approx \frac{T_m}{\rho_s q_m} \frac{{\Gamma}}{d_{\rm min}}$, where $\rho_s$ and $q_m$ are the density and heat of fusion of ice.    As we shall see, in the general case there is no strict separation of the thickness dependence of $\gamma_{ss}(d)$ because $F_{\rm short}(d)$ can be quite long range, which makes the situation unique. 

Our theoretical approach employs the full frequency dependent
dispersion (van der Waals) force contributions to the long ranged
interaction (e.g., \cite{ElbaumSchick, Wilen}) and Poisson-Boltzmann
theory for screened Coulomb interactions.  The same  basic physical
phenomena form the basis of Derjaguin-Landau-Verwey-Overbeek (DLVO)
theory  \cite{DLVO, Israelachvili, Edwards}.  The principal
difference between this approach and the {\it modified} DLVO theory
used previously to study surface and interfacial melting
\cite{Wett99a} is that the van der Waals interactions are calculated
rigorously rather than using the phenomenological treatment of
relevance in that case in which dispersion forces could be treated in
the nonretarded limit.  

There are two motivations for taking a more complex approach.  First, when the
{\it only interactions} in the system are non-retarded dispersion forces, or van der Waals forces, 
to the total excess surface free energy per unit area is most
commonly denoted by ${F_{vdW}}(d)~=~-~{{{A_H}}/{12~\pi~{d^2}}}$,
where ${A_H}$ is the Hamaker constant (e.g., \cite{Israelachvili}).
For  identical substrates (e.g., ice/water/ice) separated by distance $d$ the
Hamaker constant is positive, producing an attraction and consequently, as has been pointed out by other authors \cite{schick,lipowsky}, grain boundary melting under
dispersion forces  must be incomplete.  If however the
substrates are separated by an electrolyte solution, they may be held
at bay by repulsive screened Coulomb interactions.   Second, for
dissimilar materials, the van der Waals contribution can be both
attractive and repulsive and one can observe oscillations in the
force versus distance/film thickness curve  leading to the
possibility of the system being trapped in a local rather than global
minimum \cite{ElbaumSchick, Wilen,  BarZiv, Fenzl}.  Such is the case
of the surface and interfacial melting of ice
\cite{ElbaumSchick,Wilen}.  We do not expect this particular
behavior in the case of the grain boundary under the influence of dispersion forces alone.  However, 
when combined with repulsive screened Coulomb interactions, the effective interfacial free energy does indeed exhibit behavior that is analogous to the dispersion force contribution to the interfacial melting of ice.  For example, as discussed in general terms above, depending on the electrolyte concentration, we can observe a global minimum at infinite, finite, or zero film thickness, but with a maximum at intermediate film thickness and a positive monotonically decreasing free energy at large film thickness.  These phenomena, in part associated with retardation, thereby heightening our appreciation of the
importance of a quantitative treatment of the long ranged forces in
the system requiring a more complete calculation of ${F_{vdW}}(d)$.

The calculation of frequency
dependent dispersion force contributions to grain boundary melting
are described in section \ref{DLPSection}, but  it is prudent to
first remind ourselves of the general understanding of ice surfaces
under the influence of such interactions.  Generally speaking, when
dispersion forces dominate, the wetting of {\em any} ice surface by
water at temperatures below $T_m$ will be facilitated  when the
polarizability of the water lies between that of the ice and the
other material be it a gaseous phase, a chemically inert solid or
ice.    A particular novelty of the system, first pointed out by
Elbaum and Schick \cite{ElbaumSchick} in their study of the surface
melting of ice, is that the appropriately--transformed polarizability
of ice is greater than that of water at frequencies higher than
approximately $2 \times 10^{16}$ rad s$^{-~1}$, whereas it is smaller
at lower frequencies.  Thus, so long as the surface melted layer of
water is thin, the polarizabilities at all frequencies contribute
additively to $F_{vdW}(d)$, whereas upon thickening {\em
retardation}--the attenuation of the interaction due to the finite
speed of light--reduces the high frequency contributions and favors
those in which the polarizability of water dominates over that of the
ice.   Hence, with regard to the dispersion force
contribution to the premelting of an ice surface, we expect the
process to be complete when the third phase in the system possesses a
low frequency  polarizability greater than that of water, and
unattainable if the low frequency  polarizability  is less than that
of water. Many, if not most, solids have dielectric properties that lead to complete
interfacial premelting whereas the vapor phase of ice does not.

The great sensitivity of surface melting experiments to the presence of impurities led one of us
\cite{Wett99a} to introduce electrolytes into a model that did
not include retardation.   Here, as in that case, we consider the ions that underlie the repulsive screened Coulomb interactions in the canonical ensemble and the entire ice/solution
system is treated in the grand canonical ensemble \cite{Wett99a}.
Hence, because the ions are insoluble in the solid phase and the
Debye length characterizing the range of the interactions is
inversely proportional to the square root of the ion density, then
the range and strength of this contribution to grain boundary melting
is sensitive to the impurity level.  Therefore, it is required that
we properly capture the competition between repulsive screened
Coulomb interactions and attractive dispersion forces at very long
range, which is done through the use of Debye-H\"uckel theory in the
appropriate ensemble and  the complete theory of Dyzaloshinskii,
Lifshitz and Pitaevskii,  \cite{DLP}.  We now describe the various contributions to
the excess surface energy in turn, the total free energy of the system, and the associated 
predictions. 

\subsection{\label{Ions}Ionic Force Effects}

The screening of surface charge is described by the {\em
Poisson-Boltzmann (\rm PB) equation} for the electrostatic potential $\psi$
created by the distribution of ions of number density ${n}(z)$
throughout the film. For a monovalent electrolyte, and a surface
potential $\psi_s$ less than 25 mV,  the Poisson-Boltzmann
equation can be linearized in the {\em Debye-H{\"u}ckel}
limit which yields
\begin{equation}
\label{DH}
\psi(z) =  \psi_s e^{-\kappa z}, \qquad \mbox{where} \qquad
\kappa^{-1} = \left({{\epsilon \epsilon_o {k_b}T}\over{e^2{n}_{b}}}\right)^{1/2}
\end{equation}
is the Debye length, which captures the characteristic fall off of
the ion field in the direction $z$ normal to the surface.  The bulk
ion density is ${n}_{b}$,  $\epsilon$ is the dielectric constant of
the film, $\epsilon_o$ is the free space permittivity, $e$ is the
elementary charge, and ${k_b}T$ is the usual thermal energy.   The
repulsive force between two charged surfaces originates in the
restriction of
the configurational entropy of the ions as the surfaces are brought
closer.  The resulting excess interfacial free energy per unit area across a film of
thickness $d$ is written
\begin{equation}
\label{FDH}
F_{DH}(d)\approx {{2 {q_s}^2}\over{\kappa \epsilon \epsilon_o}}e^{-\kappa d}~,
\end{equation}
where $q_s$ is the surface charge density, discussed in more detail in section \ref{Total}.    
In the classical context of DLVO theory, the Debye length is a constant, and equation (\ref{FDH}) can be considered in the same sense as  $F_{\rm short}(d)$ above.  At this stage we simply note that it describes the electrostatic contribution to total excess
interfacial free energy of relevance to the grain boundary melting of
ice in the presence of electrolytes.

We point out several important modifications to PB theory beyond the obvious simplifications
associated with the linearization, or weak coupling limit (e.g., \cite{NetzA}),  that led to equation
(\ref{DH}).  First, as a mean field theory, ion-ion correlations and discrete ion--solvent interactions are not captured and the solvent is treated as a continuous medium \cite{AndelmanA}.  Second, 
there are dispersion forces between ions and the surfaces that bound them, which can be of much longer range than Debye-H{\"u}ckel theory would predict although not of significant magnitude at a range beyond about 3 nm where retardation becomes important \cite{Edwards}.   Such ion dispersion forces can influence the {\it static} van der Waals interactions in the system significantly for high dopant levels or multivalent ions \cite{NetzB}. Third, at molecular length scales steric, or finite ion size, effects are important \cite{AndelmanB}.  Although all of these phenomena may be of importance
in a limited regime of the predictions presented here, we do not
expect them to be responsible for a qualitative difference the principal features described.  Moreover, other equally distinct effects can also play a role.  For
example, in the case of the ion dispersion phenomena, the fact that
the chemical potential of the ions in ice is effectively zero results
in the film concentrating or diluting as a function of the departure
from bulk coexistence leads to a film thickness dependence of the
Debye-length.  Hence, we find a very long range repulsive interaction
with an effectively algebraic structure.  As for steric influences,
at a grain boundary other proximity effects associated with the
crystallographic mismatch will be come equally important at short
range.  A theory that systematically includes all such effects on
equal footing is still unwarranted by the state of the experimental
landscape, but we are compelled to note for the reader that we are
acutely aware of all such complications and limitations enumerated in the detailed studies cited in this paragraph and in books on the topic \cite{Israelachvili}, \cite{Ninham}.

\subsection{\label{DLPSection}Dispersion Force Effects}

The frequency dependent dispersion force, or van der Waals, contribution to the
free energy of a surface consisting of a pure
liquid water layer of thickness $d$ between two slabs of bulk ice can
be obtained from the theory of Dzyaloshinskii, Lifshitz, and
Pitaevskii \cite{DLP} (DLP).  The result of this theory is an
integral expression for the interfacial free energy per unit area, $F_{vdW}(d)$,  in
terms of the frequency-dependent dielectric
polarizabilities of the ice($s$) and the water($\ell$). 
In the sense described in section \ref{Background}, were grain boundary melting under
dispersion forces alone to be considered, then $\gamma_{ss}(d)= 2\gamma_{s{\ell}} +   F_{\rm long}(d) \equiv 2\gamma_{s{\ell}} + F_{vdW}(d)$.   Were complete grain boundary melting to be expected, and it is not, it would be
indicated by a global minimum of $F_{vdW}(d)$ at $d \rightarrow
\infty$ so that $\gamma_{ss}=2\gamma_{s{\ell}}$ at bulk coexistence. Incomplete grain boundary melting under dispersion
forces is indicated by a minimum at finite $d$, with $F_{vdW}(d)$
negative there.  

As described in \ref{Background}, retarded potential effects are known to influence the qualitative nature of melting in ice.    For example, the surface melting of ice has
been observed to be incomplete due to retarded potential effects, whereas slight impurity doping leads to complete surface melting \cite{Elbaum93, ElbaumSchick}.  Because the neglect of retardation can lead to qualitatively and quantitatively different predictions we append a brief extended discussion of the matter. 

In order to make this development reasonably self contained we present the DLP expression for
$F_{vdW}(d)$ despite the fact that is has appeared in a variety of forms \cite{DLP,ElbaumSchick,BarZiv,Wilen}, and with
sympathy to the girth of the literature we attempt the utmost brevity.  The frequency dependent dielectric response of the ice/water/ice system is
described by
\begin{widetext}
\begin{equation}
\label{integral}
F_{vdW}(d) = \frac{kT}{8\pi{d^2}}{\sum_{n=0}^{\infty}}^{'}
\int_{r_n}^{\infty} dx~x
\left( {\ell}n \left[
1 - \left(\frac{x  - x_{s}}
{x + x_{s}}\right)^2 e^{-~x}
\right]
+ {\ell}n \left[
1 -  \left(\frac{{{\epsilon}_s}x - {{\epsilon}_\ell}x_{s}}
{{{\epsilon}_s}x + {{\epsilon}_\ell}x_{s} }\right)^2 e^{-~x}
\right]
\right),
\end{equation}
\end{widetext}
where
\begin{equation}
x_s = {\left[ x^2 - {r_n}^2 \left(1 - \frac{{\epsilon}_s}{{\epsilon}_\ell}
\right)
\right]}^{1/2},
\end{equation}
and the material $(s,\ell)$ dielectric functions, corresponding to
ice ($\epsilon_s$) and water ($\epsilon_\ell$) are evaluated at the
sequence of imaginary frequencies $i{\xi}_n =
i({2{\pi}kT}/{\hbar})n$. The prime on the sum, indicates that the
$n=0$ term is weighted by $1/2$. The lower limit of integration is
$r_n = 2d({\epsilon}_{\ell})^{1/2}{{\xi}_n}/c$, and $k,\hbar$ and $c$
have their usual meaning. The dielectric function required in the
integral, $\epsilon(i\xi)$, is the analytic continuation of the
material dielectric function $\epsilon(\omega)$ to imaginary
frequencies. Lacking complete spectra for ice and water, as can be
obtained in high melting temperature materials (e.g., \cite{Roger}) we must
generate
the function by fitting the dielectric response of the
material to a damped-oscillator model of the form,
\begin{equation}
\label{oscillator}
\epsilon(\omega)=1+\sum_j {f_j\over e_j^2-i\hbar\omega g_j-( \hbar \omega )^2}
\end{equation}
\noindent where $e_j$, $f_j$ and $g_j$ are fitting
parameters\cite{Parsegian}. Each term in the sum corresponds to an
absorption band of frequency, width and oscillator strength, $e_j$,
$g_j$ and $f_j$, respectively. Substitution of $i\xi$ for $\omega$
gives $\epsilon(i\xi)$, a well behaved, monotonically decreasing,
real function of $\xi$.  

We note that equation (\ref{integral}) assumes implicitly that $\epsilon_s$ is isotropic. Although the
effect of the crystal orientation has been studied theoretically \cite{Tosatti}, no such polarization data was available for the present calculation.  It has not been measured over the full frequency range, but judging from known values at optical frequencies \cite{hobbs}, it is estimated to be small, and hence we neglect the small orientational dependence of the polarizabilities.  Therefore,  we
treat the ice crystals as continuum dielectrics, with the surface crystal structure embodied by the surface charge density as described in section \ref{Ions}.

\subsection{\label{Total} Total Free Energy}

The stable existence of an impure grain boundary film at temperatures below $T_m$ is investigated by minimization of the total free energy of the system. The distinction between the canonical DLVO setting, say in colloid science, and surface, interfacial or grain boundary melting is that in the former case the Debye length is constant for all values of the film thickness whereas this is not the circumstance in the latter case \cite{Wett99a}.  We imagine doping a pure film with a monovalent electrolyte such as $NaCl$.  Depending on the nature (low angle or twist) and magnitude
of the mismatch between the crystals flanking the grain boundary, the
solid--liquid interface will have a particular surface charge density
$q_s$. We realize that each solid--liquid interface can, in
principle, have a different value of $q_s$, and we embody
the surface crystal structure through the surface charge density,  but for the sake of
simplicity and pedagogy we assume that both are equal.   The
surface charge is screened by the ions in the film, creating a repulsive interaction which competes with attractive long ranged van der Waals interactions and this embodies the excess surface energy.  
 
We deposit $N_i$ moles per unit area of a monovalent electrolyte in the grain boundary.  The chemical potential of the electrolyte in ice is essentially zero \cite{hobbs}, and because we are well below the solubility limit the ions remain in the film with a concentration that is inversely proportional to film thickness.  When the temperature decreases the film thins, the ion concentration increases and hence the Debye length decreases (see equation (\ref{DH})).   At very low film thicknesses and very high ion concentrations the limitations of PB theory, discussed in section \ref{Ions} come to the fore.
Within these approximations the behavior of the {\em total} free energy of the {\em system} determines the nature of grain boundary melting.  For this we combine bulk and surface terms  as
\begin{eqnarray}
G_T(T,P,d,N_i) =\rho_{\ell}\mu_{\ell}(T,P)d + \mu_{i}(T,P){N_i}
\nonumber\\
+ RT{N_i}~{\ell}n~{{N_i}\over{\rho_{\ell} d}} +
\mu_{s}(T,P)N_s + \gamma_{ss}(d),
\end{eqnarray}
where $\gamma_{ss}(d) = 2\gamma_{s{\ell}} + F_{DH}(d) + F_{vdW}(d)$ is the total effective interfacial free energy combining equations (\ref{FDH}) and (\ref{integral}).  The molar density, chemical potential per mole and the number of moles per unit area of the solvent are $\rho_{\ell}$, $\mu_{\ell}$ and $N_{\ell}$=$\rho_{\ell} d$ respectively and the chemical potential
of the impurity, the solid and the number of moles of the latter are $\mu_{i}$, $\mu_{s}$  and $N_{s}$.  We write the mixing entropy term, $RT{N_i}~{\ell}n~{{N_i}\over{\rho_{\ell} d}}$, where $R$ is the gas constant, in the dilute limit of ideal solution theory although we could simply replace the molar ratio by the activity coefficient for the impurity in the solvent.  These complications do not change the principal regimes exhibited by the system.   

The total free energy is minimized with respect to film thickness at fixed temperature and dopant concentration and the integral in Eq. (\ref{integral}) and the dielectric fits for ice
and water compiled from reference \cite{dielectricdata}, were calculated in the same manner as was done previously \cite{Wilen},\cite{ElbaumSchick}.   The Gauss-Legendre
quadrature integration scheme was tested against these calculations
for the case of surface melting in the absence of impurities.

\section{\label{Results} Results}

A representation of what we see overall as we vary the parameters is shown in the figure. In a given calculation, we ascribe a surface charge density to both solid/film interfaces, with, for example $q_s$ = 0.01 C $m^{-2}$ corresponding to approximately four elementary charges per 1000 molecular sites.  This particular case is rather high from the perspective of counting missing bonds
at the interface or ionization defects, but conservative
from the perspective of charge adsorption. It is useful to begin in the upper left hand
corner of figure 1(a) at high temperature. Consider the most dramatic
curve first; the solid line which has the lowest dopant level of 6 $\times$ 10$^{-5} \mu$M$m^{-2}$.
Due to the fact that in this case the Debye length is extremely long and the repulsive Coulomb interaction is longer ranged than the attractive van der Waals interaction, a very thick
film of liquid exists at the ice grain boundary.  As the temperature is lowered, the efficiency with which ice rejects electrolytes causes the film to be enriched with ions, and hence the Debye length,
and the range of the Coulomb interaction, decreases until such point that the attractive tail of the van der
Waals force begins to come into its own. As the temperature decreases
a point is reached where abruptly the van der Waals interaction
dominates and drives a discontinous transition to zero film
thickness.  The transition moves to higher temperatures as the surface charge density and dopant levels decrease  thereby showing the expected behavior for melting driven solely by dispersion forces, that is, no grain boundary melting.  An increase in the initial dopant level reduces the Debye
length and the strength {\em and range} of the screened Coulomb interaction.  Hence,  the plateau where the van der Waals and screened Coulomb interactions compete shifts to lower film thicknesses.
If the temperature and/or the dopant level is
sufficiently high, the bulk free energy dominates the
total free energy and, as found previously \cite{Wett99a}, a
surface version of Raoult's law is seen; $d\propto(T_m - T)^{-1}$.   Finally, the largest dopant levels
lead to the bulk free energy dominating through the entire range,
and we predict a measurable film thickness in an easily accessible and
controllable experimental range of temperature from 0 to - 10$^0$C.  Although this
latter behavior was observed in studies of interfacial melting using atomic
force microscopy \cite{Butt},  there exist no observations on ice
grain boundaries. We emphasize that at fixed temperature the film thickness is not a monotonically increasing function of ion concentration {\em until} a threshold level is surpassed.  Thus, there is a transition from melting dominated by dispersion and ionic force interactions to that dominated by the more commonly considered colligative effects.  An exploration of variations in dopant levels and surface charge density reveals the same qualitative behavior.

\begin{figure}[h] 
	\begin{center}
		\includegraphics[width=0.25\textwidth]{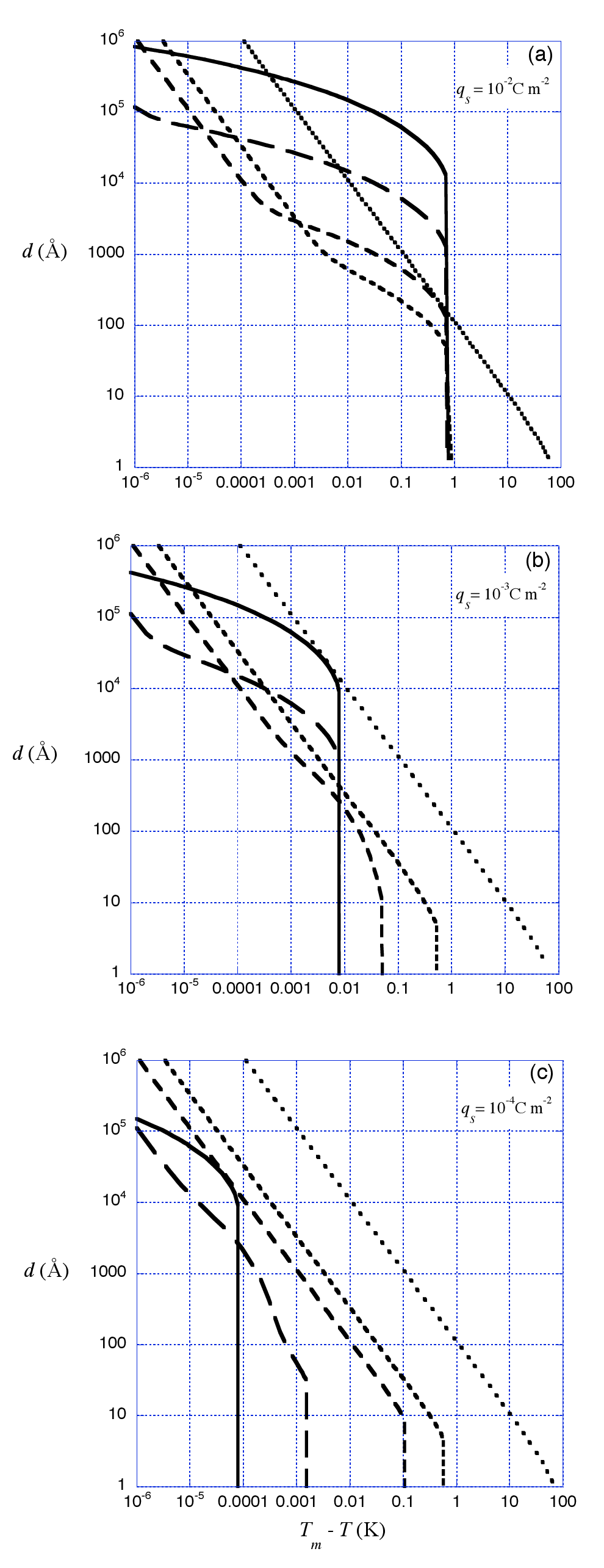}
	\end{center}
	\caption{\label{one} Film thickness (\AA) versus undercooling (Kelvin) for various monovalent electrolyte ($NaCl$) concentrations at an ice grain boundary with a solid/film surface charge density $q_s$ of (a) 0.01 C $m^{-2}$.  (b) 0.001 C $m^{-2}$,  and (c) 0.0001 C $m^{-2}$. 
In (a)-(c) ${N_i}$ = 6 $\times$ 10$^{-5}$ (solid line), 
6 $\times$ 10$^{-4}$ (long dashed line), 6 $\times$ 10$^{-3}$ (medium dashed line), 1.8 $\times$ 10$^{-2}$ (small dashed line) and 6 (dotted line) $\mu$M$m^{-2}$.    At the lower dopant levels where the Debye length is long we observe an abrupt or discontinuous onset of grain boundary melting. The transition moves to higher temperatures as the surface charge density and dopant levels decrease thereby showing the expected behavior for melting driven solely by dispersion forces.  Note, for example in (a), that at $T_m - T$ = 0.01 K,  the film thickness is not a monotonically increasing function of ion concentration.  Indeed, the film thickness {\em decreases} with increasing concentration {\em until} a threshold level is surpassed.  Hence, there is a transition from melting dominated by dispersion and ionic force interactions to that dominated by the more commonly considered colligative effects. In (b) and (c) we see that for low surface-charge densities the Coulomb interaction is weak, resulting in a much less-pronounced plateau, or complete absence thereof, except for the lowest dopant concentrations.}
\end{figure}

\section{Discussion}

One effect not considered in the current theoretical framework that warrants discussion
is the possible dependence of $\gamma_{s\ell}$ on impurities adsorbed at the
interface.  The low solubility of electrolytes in ice leads to an enhanced interfacial adsorption and will  modify the solid/film interfacial free energy, $\gamma_{s\ell}$,  in a manner that depends
on, among other things, the magnitude of the surface charge density and
the bulk electrolyte concentration.   A {\em decrease} in $\gamma_{s\ell}$
will {\em enhance} grain boundary melting relative to the pure case, were there not a similar decrease in the dry grain boundary energy.  Although it is well known that an image charge effect leads to an {\em increase}  in the liquid/vapor interfacial free energy with electrolyte concentration, there is no quantitative measure of the phenomenon at the solid/liquid interface wherein the intrinsic ionization defects in the solid phase are available to screen the adsorbed ions.  Indeed, the great sensitivity of the surface melting of ice to impurities, in part responsible for the transition from incomplete to complete melting \cite{Elbaum93}, is quite strikingly demonstrated in the calculations shown here.  However, deconvolution of the precise dependencies of the constituents of the interfacial energies on $N_i$, the coupling to the ionic forces in the solution within the PB framework, and the additional influences of ion-ion correlations, ion-solvent and ion-dispersion interactions, and steric effects described above, is beyond the scope of experimental evidence.  

As described above, there are two principal ways in which grain boundary melting can be observed;  scrutiny of the dihedral angle subtended by a grain boundary groove, or direct probing of the liquid film thickness at the boundary as a function of temperature.  In principal, the dramatic nature of our predictions and their thermodynamic accessibility renders them amenable to an experimental search.  In the experimental scenarios that we envisage, the grain boundary of a bicrystal is in contact with the bulk solution at a classical grain boundary groove and hence, it is theoretically possible to examine both the dihedral angle {\it and} the grain boundary thickness.  However, the ensemble operative in our predictions leads to a particularly impurity sensitive film thickness dependence of the Debye length; Equation (\ref{DH}) leads to  $\kappa = c~\sqrt{{N_i}/d}$, where $c$ = 7.237 $\times$ 10$^7$ m$^{{1/2}}$ mol$^{-{1/2}}$ is a constant.  If however, contact of the grain boundary film with a bulk reservoir of ions allows compositional equilibration between the film and the bulk, then the Debye length will not be a function of the film thickness and some qualitative aspects of the predictions will differ.   We expect compositional equilibration for the thickest films, but the presence of a substrate field in the case of the thinner films may not facilitate such homogenization.  This question is the subject of ongoing work.  

There are different methods that probe different aspects of premelted liquid water.  For example, infrared spectroscopy averages the premelted liquid in a polycrystalline sample \cite{Vlad}, introduction of a force microscope tip of another material provides and indirect probe \cite{Butt}, \cite{Fain}, bright X-ray reflectivity discerns short range structural properties \cite{Dosch} and optical methods (e.g.,  \cite{Elbaum93}, \cite{Optical}) distinguish between liquid water and ice.  Regardless of the approach, our calculations indicate the preference of systematic doping over attempts to prepare a completely clean system. 

In summary, we have studied the effect of impurities on the grain boundary melting of ice using an extension of Derjaguin-Landau-Verwey-Overbeek
theory, in which we include retarded potential effects in a
calculation of the full frequency dependent van der Waals and
Coulombic interactions within a grain boundary.   We find that, depending
on the amount of impurity and the surface charge density,  as
temperature decreases, attractive dispersion force
interactions effectively compete with repulsive screened Coulomb interactions producing grain boundary melting through a discontinuous transition.   At sufficiently high dopant
concentrations the classical solutal effect dominates the melting
behavior.  The effect is within the scope of experimental accessibility.

\begin{acknowledgments}
Conversations and criticism from J.G. Dash,  J. Neufeld, A.W. Rempel, E. Thomson and L.A. Wilen are greatly appreciated.  We wish to acknowledge the support of the NSF-OPP9908945, the Bosack and Kruger Foundation and Yale University. 
\end{acknowledgments}

\appendix

\section{Retardation Effects}

The detailed quantitative effects of retardation mentioned in section \ref{Background} are placed in this Appendix because grain boundary melting under dispersion forces alone is incomplete.  Retardation has been clearly described previously (e.g., \cite{ElbaumSchick,BarZiv,Parsegian, Wilen}) by considering a general ice/water/{\cal X} interface in which {\cal X} denotes a ``substrate'' which may be pure water vapor.  In the limit that $\epsilon_\ell \approx \epsilon_s \approx \epsilon_X\approx 1$, one can approximate Eq. (\ref{integral}) as;
\begin{equation}
\label{approx}
F_{vdW}(d)\approx
-~\frac{kT}{8\pi{d^2}}
{\sum_{n=0}^{\infty}}^{'}
\left(\frac{{\epsilon}_s - {\epsilon}_\ell}
{{\epsilon}_s + {\epsilon}_\ell}\right)
\left(\frac{{\epsilon}_{X} - {\epsilon}_\ell}
{{\epsilon}_{X} + {\epsilon}_\ell}\right)
(1+r_n) e^{-~{r_n}}.
\end{equation}
The term $ e^{-~{r_n}}$, due to retardation, acts as a high frequency cutoff to the sum which is inversely proportional to $d$.

When the substrate is pure water vapor, $\epsilon_X$ may be taken equal to 1. In this case, it is clear that were $\epsilon_s-\epsilon_\ell < 0$ at all frequencies, $F_{vdW}(d)$ would be a monotonically 
{\em increasing} function of $d$ and the film would not grow. Conversely, if $\epsilon_s-\epsilon_\ell >0$ at all frequencies, then $F_{vdW}(d)$ would be a monotonically {\em decreasing} function and the film thickness would diverge as the melting temperature is approached.  Because $\epsilon_s -
\epsilon_\ell$ changes sign at frequency $\xi_c$, the resulting melting behavior  is intermediate between these cases. In particular, for sufficiently large $d$, where the sum is dominated
by the low frequency terms, surface melting is inhibited. When the substrate is a solid material with arbitrary dielectric properties, the results are more complicated. The function $\epsilon_X$ now depends on frequency $\xi$, and $\epsilon_\ell(i\xi)-\epsilon_X(i\xi)$ may change sign. For all of the materials studied previously \cite{Wilen}, we observed that $\epsilon_X >\epsilon_{s,\ell}$ in the frequency range $\xi < \xi_c$. Hence we found that for $d$ large enough ($\approx$30\AA)
to allow retardation to come into play, $F_{vdW}(d)$ will be a positive monotonically decreasing function of $d$, implying that the necessary condition for complete interfacial melting to occur is fulfilled. To determine whether there is another, deeper, minimum in the free energy at small film thicknesses requires implemention of the full dispersion force calculation embodied in equation (\ref{integral}).

\end{document}